\documentclass{epl}

\title{First-order microcanonical transitions \\
in finite mean-field models}
\shorttitle{First-order microcanonical transitions}
\author{Mickael Antoni\inst{1} \and Stefano Ruffo\inst{2} 
\and Alessandro Torcini\inst{3,2,1}}
\shortauthor{M. Antoni {\it et al.}}
\institute{
\inst{1} UMR-CNRS 6171 - Universit\'e d'Aix-Marseille III -
Av. Esc. Normandie-Niemen, 13397 Marseille Cedex 20, France. \\  
\inst{2} Dipartimento d'Energetica ``S. Stecco'' and CSDC 
, Universit\`a di Firenze, and INFN and INFM, 
via S. Marta 3, 50139 Firenze, Italy. \\
\inst{3} Istituto Nazionale d'Ottica Applicata, Largo E. Fermi 6, 
50125 Firenze, Italy. \\
}
\pacs{05.70.Fh}{Phase transitions: general studies}
\pacs{05.70.Ln}{Nonequilibrium and irreversible thermodynamics}
\pacs{05.40.-a}{Fluctuation phenomena, random processes, 
noise, and Brownian motion}
\begin{document}

\maketitle

\begin{abstract}
A {\it microcanonical first order transition}, connecting a
clustered to a homogeneous phase, is studied 
from both the thermodynamic and dynamical point of view 
for a $N$-body Hamiltonian system with infinite-range couplings.
In the microcanonical ensemble specific
heat can be negative, but besides that, a microcanonical first order
transition displays a temperature discontinuity as the energy is varied
continuously (a dual phenomenon to the latent heat in the canonical
ensemble). In the transition region, the entropy per particle exhibits,
as a function of the order parameter, two relative maxima separated by
a minimum. The relaxation of the {\it metastable} state is
shown to be ruled by an activation process induced by intrinsic finite
$N$ fluctuations. In particular, numerical evidences are given that the 
escape time diverges exponentially with $N$, with a growth rate given 
by the entropy barrier.
\end{abstract}

\section{Introduction}

There has been recently a renewed interest for systems with long-range 
interactions ~\cite{leshouches}. Phase transitions from clustered to 
homogeneous phases, occurring in simple models of globally coupled particles, 
have been analysed within different statistical ensembles~\cite{gross_book,leshouches}.
It has been shown that, near transitions that are of the first order in the
canonical ensemble, ensembles are inequivalent~\cite{julien,cohen}. 
In the microcanonical
ensemble the phase coexistence region can display a {\it negative specific
heat}, corresponding to a convex entropy as a function of energy.
In this region, entropy can be either a continuous and infinitely
differentiable function of the energy (in the case of
a continuous microcanonical transition) or it can display a discontinuity
already in the first derivative. This case is denoted as {\it microcanonical
first order} transition and is characterized by a jump in temperature
as the energy is varied continuously~\cite{julien,ispo}\footnote{Other, more
complicated, situations have been rigorously classified in~\cite{barreth}}.

This latter case remains to be fully understood from the physical point
of view: what does the coexistence of two temperatures, at
equilibrium, for a given energy mean ? In this Letter we present a careful
analysis of the microcanonical first order transition that appears in
a toy model that describes the motion of particles in a two-dimensional
bounded domain (a torus). The model has been first introduced in Ref.~\cite{at,art},
where the analysis was limited to microcanonical continuous transitions.

In the first part of this Letter, we fully characterize the phase transition
by studying the intricate dependence of temperature on energy (the so-called
caloric curve) and the dependence of entropy on both the energy and the 
order parameter, discussing also finite $N$ effects.

In the second part we concentrate on a study of the relaxation
dynamics from a {\it metastable} state for a finite number
$N$ of particles. This process had been originally investigated by Griffiths 
{\it et al.}~\cite{gwl} within the canonical ensemble.
In this pioneering paper the authors have shown that, for a Curie-Weiss Ising 
model in an external magnetic field, the relaxation time
of the metastable states grows exponentially with $N$,
the exponential growth rate being given by the free-energy barrier per spin
\footnote{For more recent developments see \cite{binder}}.
More recently, the influence on the relaxation dynamics
of extrinsic (thermal) noise source and of the
intrinsic noise source due to finite $N$-effects has
been analysed for a $\phi^4$-model with long range 
interactions~\cite{rosti}.

We are not aware of any similar study within the microcanonical
ensemble, especially in a situation of ensemble inequivalence.
Indeed, the out of equilibrium dynamical behaviour near a
microcanonical continuous transition has been already examined
as far as the relaxation from {\it unstable} states is
concerned~\cite{ar,yama}.
The relaxation times have been shown to be typically proportional 
to some power of $N$.
In our analysis, since no extrinsic noise is present, we emphasize 
the role played by the {\it intrinsic noise source} originated by finite 
$N$ effects.
These intrinsic fluctuations induce an exponential divergence with $N$ of the relaxation 
time, with a growth rate given by the entropy barrier per particle.

\section{The model}
The model we consider is a classical $N$-body Hamiltonian system
defined on a two-dimensional periodic cell. The inter-particle potential is
infinite ranged and the particles are all identical and have unitary mass.
The Hamiltonian of the model is \( H_a = K + V_a \), where 
$K= \sum_{i=1}^{N}\left(p_{x,i}^{2}+p_{y,i}^{2}\right)/2$ 
is the kinetic energy while the potential energy reads
\begin{equation}
V_a=\frac{1}{2N}\sum^{N}_{i,j=1}\left[ 2+a-\cos
\left( x_{i}-x_{j}\right) -\cos \left( y_{i}-y_{j}\right) \right.
\left. -a\cos (x_{i}-x_{j})\cos (y_{i}-y_{j})\right]
\quad ,
\label{eqHA}
\end{equation}
where \( (x_{i},y_{i})\in ]-\pi :  \pi ] \times ]-\pi : \pi ] \) represents
the coordinates of the $i$-particle, \( (p_{x,i},p_{y,i}) \) the
conjugated momenta and $a$ is a parameter. 
The complete phase diagram of this model has been previously presented 
in Ref.~\cite{art}.
We will limit here our analysis to the parameter value $a=2$, for which the
system undergoes a first order microcanonical phase transition.
The different phases are conveniently characterized by the vector order
parameter 
\( \overrightarrow{M}_{z}=\left( \left\langle \cos (z)\right\rangle _{N},
\left\langle \sin (z)\right\rangle _{N}\right) \),
where \( z=x \) or \( y \)  and \( \left\langle \right\rangle _{N} \)
indicates the average over all the particles. It can be shown that
\( |\overrightarrow{M}_x| \approx |\overrightarrow{M}_y| = M \) and
$M$ can be thought as the {\it magnetization} of the infinite-range
Heisenberg XY Hamiltonian (\ref{eqHA}).
In the low energy $U=H_a/N$ clustered phase (CP) all particles are trapped in
a single cluster and $M \neq 0$, whereas in the high energy homogeneous
phase (HP) they are uniformly distributed in the cell and 
\( M\approx O(1/\sqrt{N}) \). At the microcanonical transition
energy, both the order parameter and temperature have a discontinuity.

We have analyzed model (\ref{eqHA}) from an analytical and a 
numerical point of view. Since the microcanonical and the canonical
variational problems define the same critical states~\cite{muk,art}, it
is convenient to first solve the model in the canonical
ensemble. Indeed, via standard saddle-point techniques, one obtains 
all absolute and relative extrema of the free energy as a function of
the order parameter~\cite{art}. These are also extrema of the entropy, 
although their stability properties are different in the two ensembles.
States that are unstable in the canonical ensemble may become stable
in the microcanonical: this is the mechanism at the origin of the
negative specific heat in the microcanocical ensemble\footnote{The first
thorough discussion of the stability of microcanonical critical states 
was performed in Ref.~\cite{katzone} for self-gravitating systems}.
From the entropy extrema, we get all the thermodynamics of the model in the
mean-field ($N\to\infty$) limit, including caloric curves 
and entropy barriers, that we discuss in the next Section.
As far as the numerics is concerned, we perform molecular dynamics 
simulations (of course at finite $N$) using an accurate fourth order symplectic
integrator\footnote{More details on the integration method can be found in
~\cite{at}}.

\begin{figure}[h]
\twoimages[width=5cm,height=5cm,angle=270]{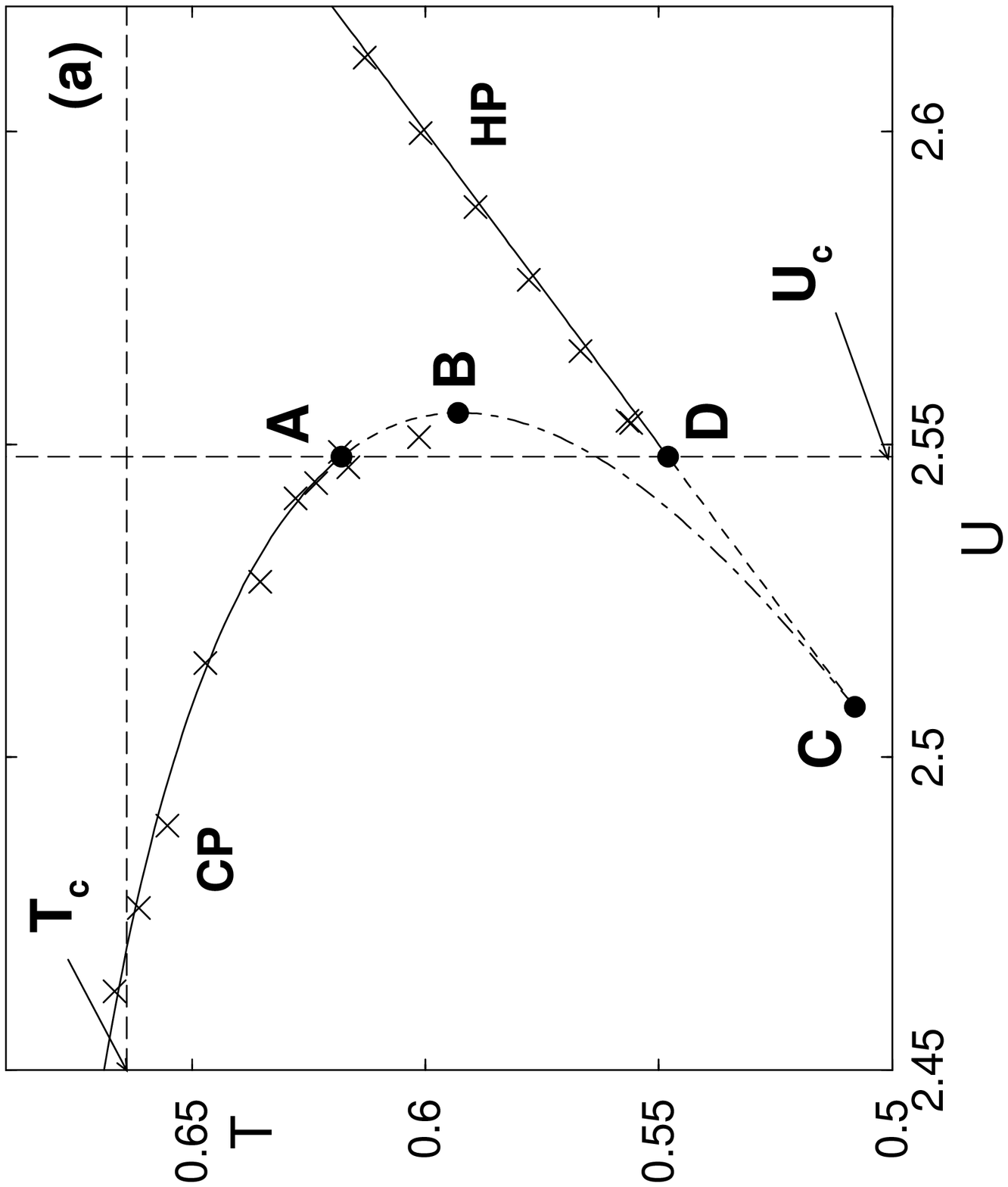}{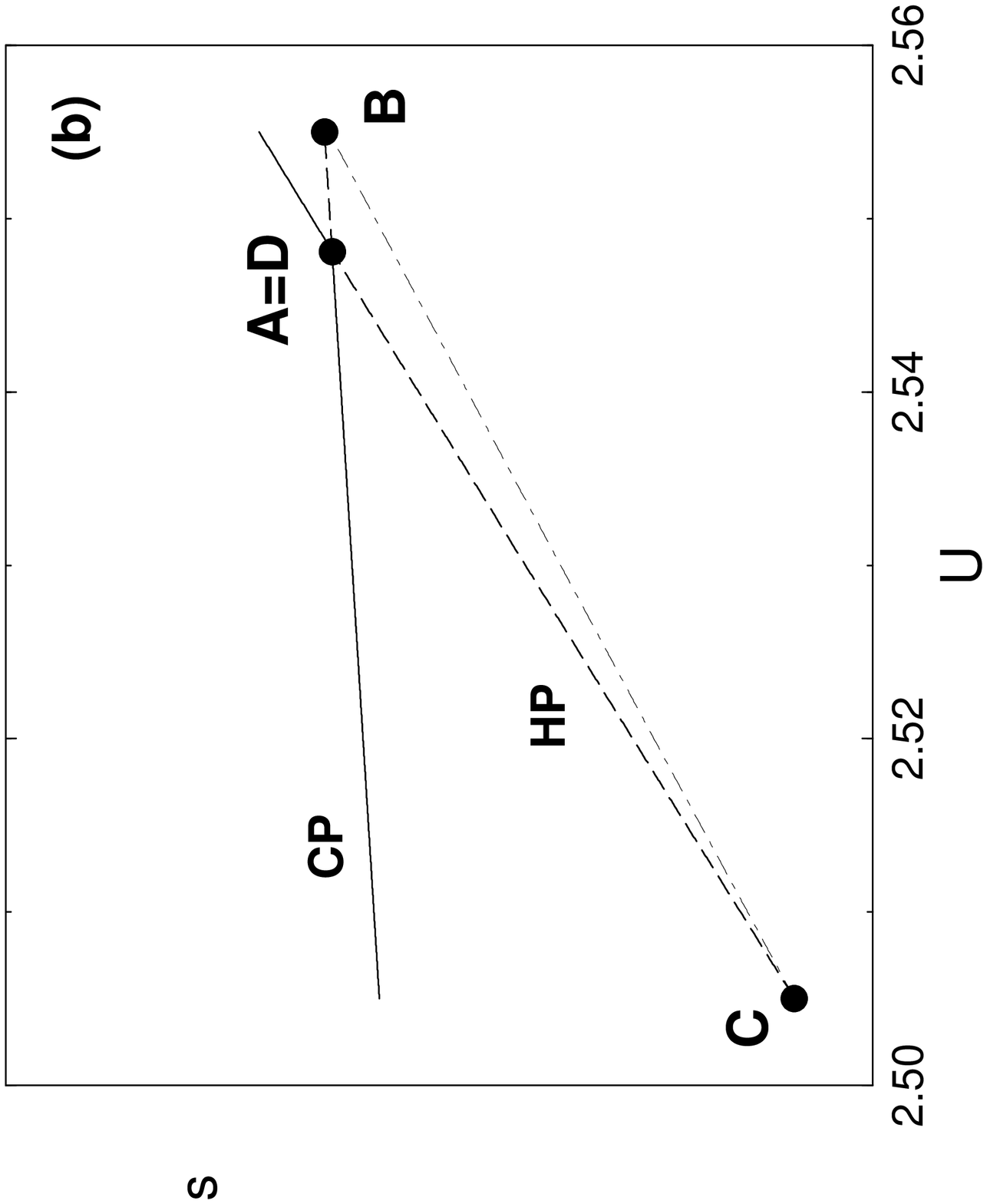}
\twoimages[width=5cm,height=5cm,angle=270]{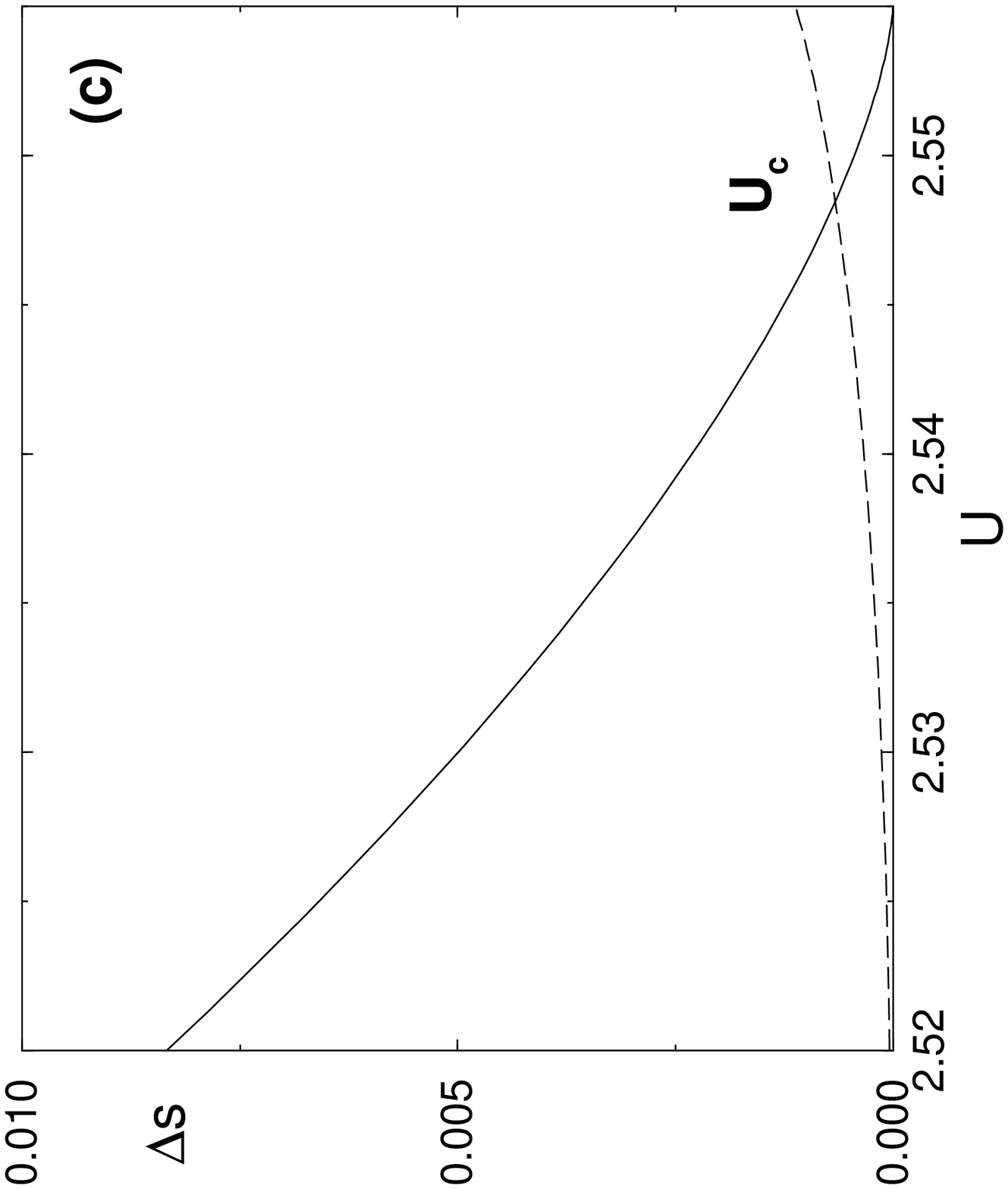}{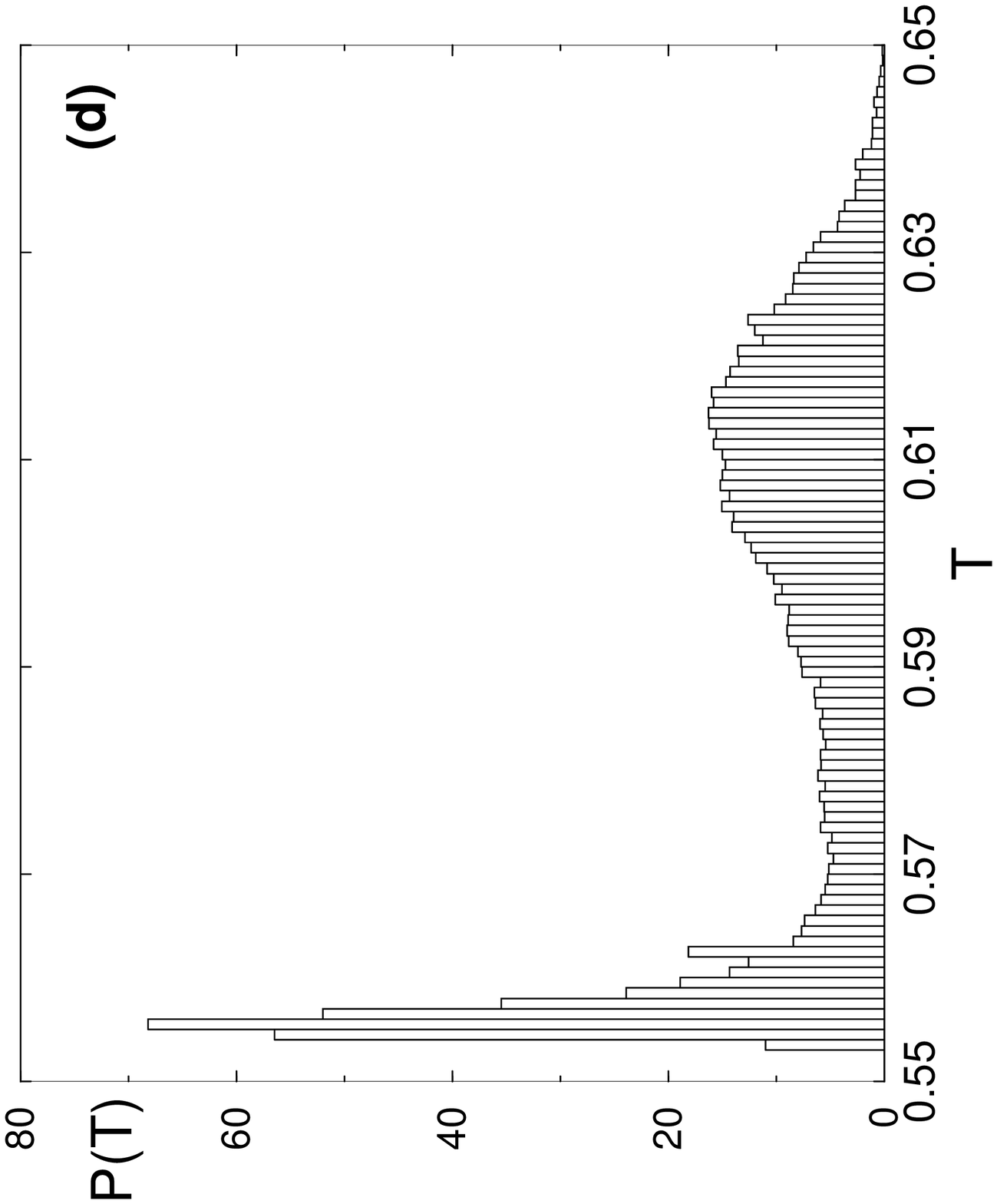}
\caption{(a) Temperature $T$ as a function of energy per particle $U$ (caloric
curve). Crosses refer to molecular dynamics simulations, where temperature is
measured by the time average of the kinetic energy, while lines indicate
mean-field analytical results. Solid lines refer to stable phases,
dashed to metastable and dot-dashed to unstable.
Numerical data have been obtained for $N=10,000$ by integrating the equations
of motion of Hamiltonian (\ref{eqHA}) for a time ranging from
$700,000$ to $900,000$ typical periods of the motion. 
The vertical (resp. horizontal) dashed line corresponds to the 
microcanonical (resp. canonical) transition
energy $U_c=2.548$ (resp. temperature $T_c=0.664$).
(b) Sketch of the entropy per particle $s$ as a function of energy $U$.
Solid lines refers to stable phases, dashed and dot-dashed 
to metastable and unstable, respectively.
(c) Entropy barriers per particle versus energy $U$.
The solid line refers to $\Delta s_{CP}$
and the dashed line to $\Delta s_{HP}$.
The lines cross at the microcanonical transition energy $U_c$.
(d) Histogram of the instantaneous temperatures close to $U_c$ ($U=2.550$) 
for $N=4,000$. Temperatures are measured by partially averaging kinetic energy
over a time interval $\Delta t=300$. (The duration of the complete run is 
$900,000$ proper periods.)}
\label{f.1}
\end{figure}

\section{Thermodynamics} 
Let us first briefly resume what is known~\cite{art}, about the phase
diagram. In the canonical ensemble, model (\ref{eqHA}) 
exhibits a continuous transition from the CP to the HP
for small $a$ values (namely for $0 \le a < 2/5$), above
the canonical tricritical point, $a=2/5$, the transition becomes
discontinuous with a finite energy jump.
In the microcanonical ensemble the transition
is continuous till the microcanonical tricritical point,
$a \sim 1.15$, is reached. For $a > 1.15$ a first order transition
with a temperature jump is observed.

We concentrate here our analysis to the specific parameter value $a=2$, 
for which the transition is discontinuous in both the canonical and
the microcanonical ensemble. In the canonical ensemble, at the transition
temperature $T_c = 0.664$, one observes an
energy jump from $U_1 = 1.896$ to $U_2 = 2.664$, corresponding
to a release of a latent heat.
In this energy range the microcanonical ensemble gives different predictions.
The CP is stable in the interval $[U_1,U_c=2.548]$, while in the interval 
$[U_c,U_2]$ the stable phase is the HP. The two phases are connected 
by a finite temperature jump at $U_c$ from $T_{CP}=0.618$ to $T_{HP}=0.548$.
These results are confirmed by the numerical simulations (see 
Fig.~\ref{f.1}(a)), although the sharp jump is
smoothed by finite $N$ effects.
The specific heat is negative from $U=2.240$ till $U_B=2.555$ (point B
in Fig.~\ref{f.1}(a)), but before reaching this energy value, at point A,
the CP becomes metastable and remains such from A to B.
At point B the specific heat vanishes and the phase turns to unstable, 
with an associated positive specific heat\footnote{Point B is a 
Poincar\'e turning point similar to that one investigated 
in Ref.~\cite{katz} for isothermal spheres.}.
The unstable curve BC (dot-dashed line in Fig.\ref{f.1}(a))
joins the CP to the HP phase ($T=U-2$). 
HP is metastable along CD in Fig.\ref{f.1}(a), 
while for $ U > U_c$ it is always stable.

The microcanonical transition can be better understood
by considering the behaviour of the entropy per particle $s$ as a function of the 
order parameter $M$. In the energy range $[U_C,U_B]$ the entropy always exhibits
two maxima separated by a minimum. The two
maxima correspond to the HP ($s_{HP}$) 
and the CP ($s_{CP}$), respectively, while the minimum
refers to an unstable clustered phase ($s_{unst}$). Plotting the
values of these extrema as a function of $U$,
three distinct branches can be drawn (see Fig. \ref{f.1} (b)):
a concave one corresponding to the $CP$, and two convex ones
corresponding to the $HP$ and to the unstable phase, respectively.
For $U < U_C$ the entropy $s=s(M)$ exhibits a single
maximum $s_{CP}$. At $U_C$, a second lower maximum $s_{HP}$ emerges
together with a minimum, via a saddle-node bifurcation. 
At higher energies  the height of $s_{HP}$ 
increases, while that of $s_{CP}$ decreases. At $U_c$ the two maxima 
reach the same height: this signals the first order transition
energy. This is illustrated in Fig. \ref{f.1}(c),
where $U_c$ is identified by the intersection
of the barrier heights $\Delta s_{CP} = s_{CP} - s_{unst}$
and $\Delta s_{HP} = s_{HP} - s_{unst}$.
For $U > U_c$ the HP maximum prevails and the CP phase becomes metastable.
At $U_B$ the entropy minimum and the lower 
maximum corresponding to the CP merge and disappear via 
an inverse saddle-node bifurcation.
Above such energy only one maximum in the entropy is
present and it is associated to the HP.

Another strong indication that the transition
is first-order comes from the fact that near
the transition energy an intermittent behaviour is observed:
the system jumps erratically from the CP to the HP and back~\footnote{
In the mean-field limit this intermittency is peculiar of the transition
energy $U_c$}.
In Fig.~\ref{f.1}(d) the histogram of the instantaneous temperatures 
is reported for a system of $N=4,000$ particles at an energy, $U=2.550$,
close to $U_c$. Two peaks are present in the histogram: one is associated to the
HP ($T \approx T_{HP} = 0.550$) and the other to the CP ($T\approx T_{CP} = 0.614$).
Hence, there is no mystery in the coexistence of two temperatures
in equilibrium: temperature, in the microcanonical ensemble, is a derived
quantity and is subject to fluctuations. The bimodal shape of the
temperature histogram reflects the analogous bimodality of the order
parameter fluctuations, which is a consequence of the first order nature
of the phase transition. Similarly, in a canonical simulation, one
would observe a bimodal energy distribution, with peaks at $U_1$
and $U_2$.

\begin{figure}[h]
\twoimages[width=5cm,height=5cm,angle=270]{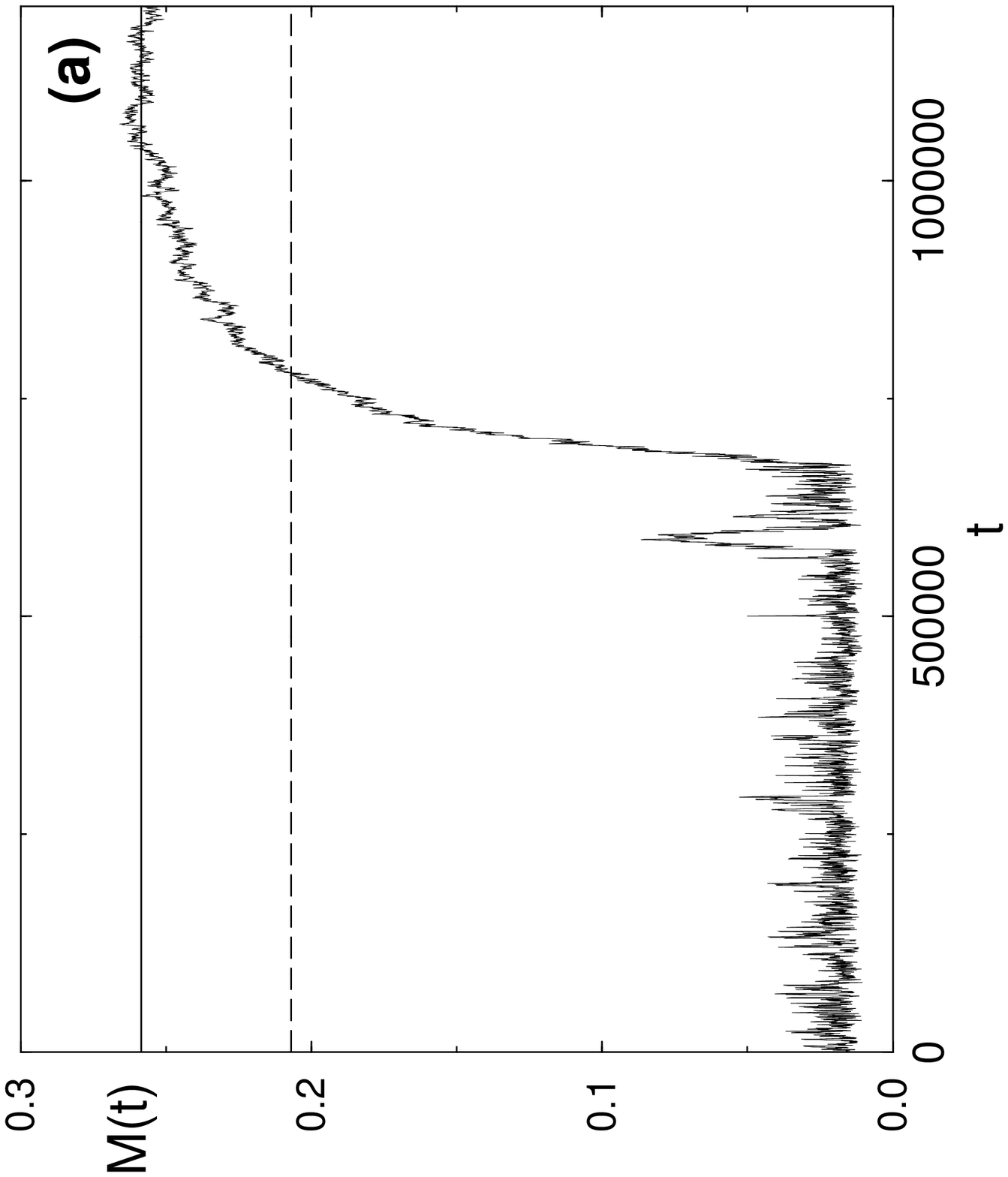}{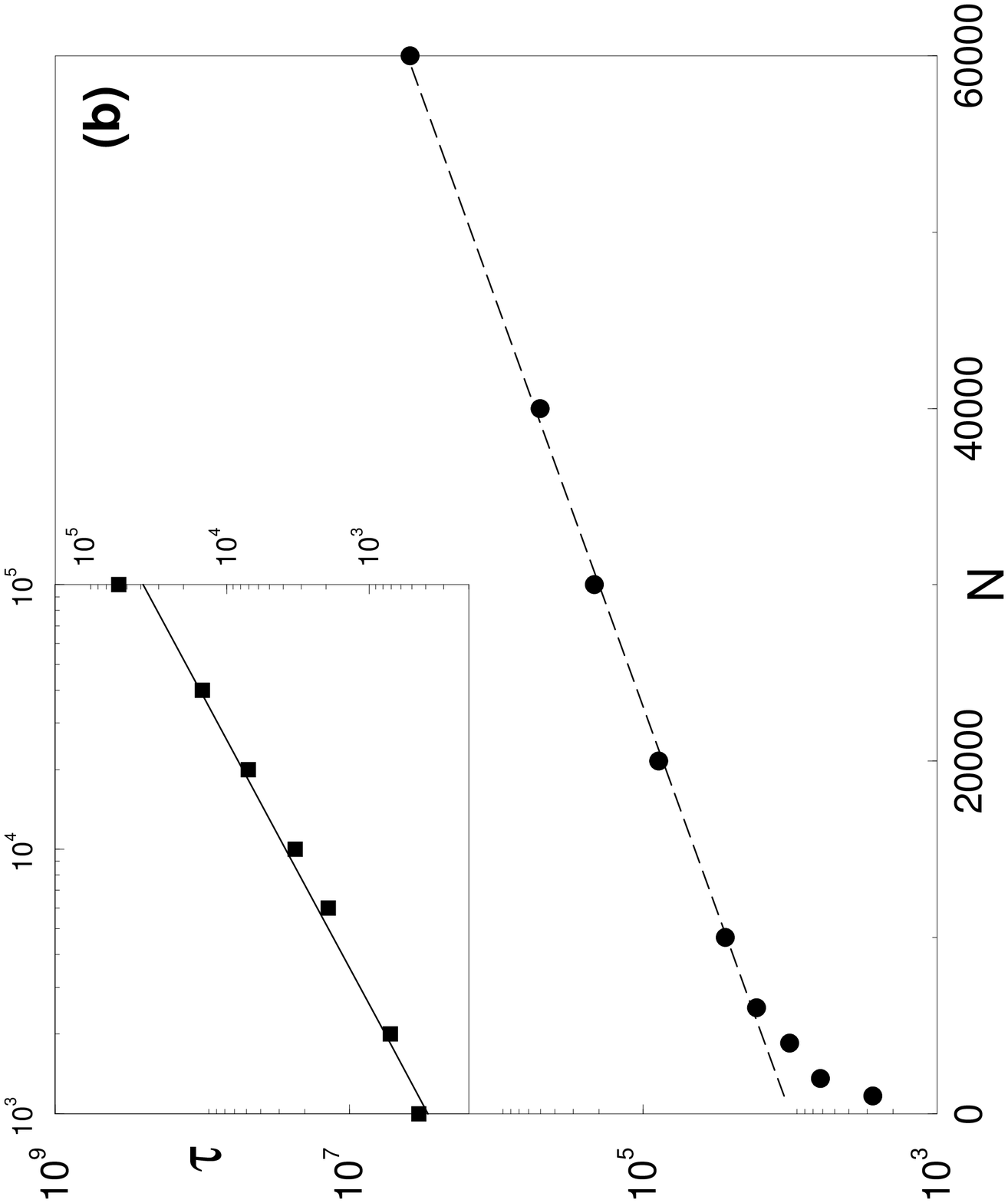}
\caption{(a) Istantaneous magnetization $M(t)$ versus time.
$M(t)$ has been obtained by averaging over
short time windows of duration $300$ and over $20$
initial conditions. The asymptotic
magnetization value is $M=0.2586$ (solid line) and 
the dashed line indicates the threshold value used
for the determination of the lifetime (i.e. 0.207).
The data refer to $U=2.530$ and $N=4,000$. (b) 
Lifetimes $\tau$ of the metastable state HP
as a function of $N$ in
a log-lin scale. Circles refer to numerical data at
$U=2.530$, the dashed line to the best fit obtained
in the interval $ 6,000 < N < 60,000$ (see Table~\ref{t.1}).
In the inset we report numerical results (squares) for
$U=2.40$ in a log-log scale, the solid line
is a linear fit to the data with slope one.
The data have been obtained by averaging over $10$ to
$1,000$ initial conditions.
}
\label{f.3}
\end{figure}

\section{Relaxation Dynamics}

We are interested in characterizing the differences in 
the relaxation dynamics from an unstable state towards a 
stable one, with respect to an
initial metastable state. As detailed above, an unstable 
state is an entropy minimum, while a metastable state
is a relative maximum, separated from the stable state by
an entropy barrier. 

It is reasonable to conjecture
that in the first case the relaxation process is dominated by 
a ``diffusive like'' behaviour, while in the second
situation the {\it activation} across an entropy barrier
should play a fundamental role.
At variance with canonical ensemble dynamics~\cite{rosti}, the activation 
process can be induced only by finite $N$ fluctuations, since there
is no coupling to an external bath.
Hence, we expect that the relaxation time from an unstable state 
will typically increase with a power law in the particle number, 
the relaxation process being dominated by the {\it granularity} of the 
system, that drives it to equilibrium through collisional effects
of the type encountered in self-gravitating systems~\cite{binney}).
Instead, for a metastable state, we should find a noticeably
different behaviour. Let us consider the situation where the CP is
stable, while the HP is metastable. By following the standard fluctuation
theory~\cite{landau}, we expect that the probability to observe a 
given value of the magnetization in the interval $[M:M+dM]$ is given by
\begin{equation}
w(M) \enskip dM = {\rm const.} \enskip {\rm e}^{[S(M) - S_{HP}]} \enskip 
dM
\quad = (2 \pi <M^2>)^{-\frac{1}{2}} 
\enskip {\rm e}^{-\frac{M^2}{2 <M^2>}}  \enskip dM
\label{prob}
\end{equation}
where $S(M) = N s(M)$ is the entropy as a function of the order
parameter, $S_{HP} = N s_{HP}$ is the entropy of the HP (for which $M=0$)
and $<M^2> \sim O(1/N)$.
As soon as the magnetization reaches the value corresponding to the 
entropy minimum $M_{unst}$, the system is quickly driven towards the CP. 
Therefore, the relaxation time from a metastable state
should diverge as :
\begin{equation}
\tau \propto [w(M_{unst})]^{-1}  \propto {\rm e}^{[s_{HP}-s_{unst}] N}
\label{tau}
\end{equation}
and we will observe an activated escape process as long as
$\Delta s_{HP} \ge 1/N $.

Let us check these two distinct relaxation behaviours in
numerical experiments. In order to measures the lifetime
of a metastable (unstable) HP,
the system is initialized with zero magnetization (indeed
due to finite $N$ effects $<M> \sim 1/\sqrt{N}$)
\footnote{The initial configurations have been 
realized in two different ways, but significant
discrepancies have never been observed. In one case the 
particle positions have been chosen randomly within 
the cell and the velocities accordingly to the
Maxwell-Boltzmann distribution. In the other one an equilibrated
configuration with energy $U> U_c$ (where the HP is stable)
is taken and the velocities are rescaled to obtain the desired energy.}. 
Once the initial state is prepared, we follow its time evolution by
monitoring magnetization.
A typical behaviour is reported in Fig.~\ref{f.3}(a).
Then, we register the time $M$ needs to reach $80 \%$ of its
asymptotic value. The lifetime of the HP state
is finally determined by averaging over several different
initial conditions. 

We first examine the relaxation of a metastable state, 
therefore the system is initialized with an energy in the
range $[U_C,U_c]$. In this interval entropy exhibits an 
absolute maximum $s_{CP}$ (corresponding to the CP)
and a relative maximum $s_{HP}$
(corresponding to the HP) separated by a minimum $s_{unst}$.
If we prepare the initial state in the HP this will have
a lifetime $\tau$ that we expect to grow as $\exp{[\Delta s_{HP} N]}$.
In particular, as shown in Fig.\ref{f.3}(b) at $U=2.530$,
for $N > 6,000$ $\tau$ has indeed an exponential growth
with $N$. Moreover, as reported in Table \ref{t.1} the numerical
estimation of the growth rates $\Delta s_{num}$ at three different energies in
the interval $[U_C,U_c]$ are in agreement with the mean-field
value $\Delta s_{HP}$. As a final remark, one should notice
that for small $N$ (e.g. for $100 < N \le 6,000$ at $U=2.530$) 
the escape time grows proportionally to $N$ instead of exhibiting 
an exponential growth (as it can be seen also in Fig.\ref{f.3}(b)).
This behaviour resembles the relaxation from an unstable state
and it is due to the fact that for such small number of particles
$1/N >> \Delta s_{HP}$. Therefore the magnetization fluctuations
are so large to easily overcome the entropy barrier, that
does not play any role here.                                
In order to measure the lifetime of an unstable HP,
we take the energy $U=2.4$, for which the
zero magnetization state is unstable and the CP is stable.
At $U=2.4$ the transition times $\tau$ scale clearly with 
$N$ as shown in the inset of Fig.~\ref{f.3}(b). 
Such type of divergence was first observed in Refs.~\cite{ar}
in model (\ref{eqHA}) for $a=0$.
Recently, Yamaguchi~\cite{yama} has studied the same 
model, finding that $\tau$ diverges as $N^{1.7}$.
On the other hand Bouchet~\cite{freddy} has shown, again for $a=0$, that
the single particle self-diffusion time increases linearly with 
$N$.

\begin{table}[h]
\caption{Entropy barries $\Delta s_{HP}$, as estimated
from the mean-field microcanonical analytical results
and by direct molecular dynamics simulations.
The escape times from the homogeneous
metastable phase are evaluated by averaging over
$10$ to $1,000$ initial conditions.
The numerical values for $\Delta s$ are
estimated by fitting the escaping times as
a function of $N$,
tipically in the interval $4,000 - 60,000$.}
\label{t.1}
\begin{center}
\begin{tabular}{lcr}
$U$   &  $\Delta s_{HP}$ & $\Delta s_{num}$ \\
2.530 & 1.28$\times 10^{-4}$ &  ($1.0 \pm 0.1$)$\times 10^{-4}$\\
2.540 & 3.33$\times 10^{-4}$ &  ($3.4 \pm 0.5$)$\times 10^{-4}$ \\
2.545 & 5.00$\times 10^{-4}$ &  ($5.8 \pm 0.4$)$\times 10^{-4}$
\end{tabular}
\end{center}
\end{table}

\section{Concluding remarks}
In this Letter we have examined a first-order
microcanonical transition from a clustered to a homogeneous
phase for a $N$-particle model with infinite-range interactions.
Even in the mean-field limit canonical and microcanonical ensembles
are inequivalent. While in the canonical ensemble
the transition exhibits an energy jump at the transition temperature, 
in the microcanonical a temperature jump is observed at the 
transition energy. Moreover, specific heat is negative in a wide
energy range. We have elucidated the intricate structure of
stable, metastable and unstable phases that coexist near the
transition.

A more physical grasp is obtained by considering the entropy of 
the system as a function of the order parameter. In a large energy domain, 
we observe the coexistence of a stable and a metastable phase, 
corresponding to the absolute and relative entropy maximum,
respectively. In this energy interval the two maxima are
separated by an entropy minimum, associated to an unstable state.
We have shown that the relaxation from the metastable state
is due, for sufficiently large $N$, to an activated crossing 
of an entropic barrier and that the activation mechanism is 
induced by finite $N$ intrinsic fluctuations.
The escape time from the local entropy maximum 
grows exponentially with $N$, and the 
growth rate is given by the entropy barrier per particle.
We have further checked that, instead, the relaxation time
from an unstable state grows linearly with $N$.

These results represent a first step towards the assessment of
a more rigorous theory of the dynamical evolution of 
a microcanonical system with long-range interactions 
trapped in a metastable state. A further refinement will
require the derivation of an appropriate Fokker-Planck
equation describing out of equilibrium fluctuations.

\acknowledgments

\noindent
We acknowledge discussions with  J. Barr\'e, F. Bouchet, E.G.D. Cohen, T. Dauxois, 
D.H. Gross and D. Mukamel.
This work has been partially supported by the MIUR (Italy) under COFIN03.


\begin{thebibliography}{0}

\bibitem{leshouches}
\Book{Dynamics and Thermodynamics of Systems with Long Range Interactions}
\Editor{Dauxois T. {\it et al.}}
\Vol{602}
\Publ{Lecture Notes in Physics, Springer, Berlin}
\Year{2002}.

\bibitem{gross_book} 
\Name{Gross D.H.E.}
\Book{Microcanonical thermodynamics:Phase transitions in ``small'' systems}
\Publ{World Scientific, Singapore}
\Year{2000}.

\bibitem{julien}
\Name{Barr\'e J., Mukamel D. \and Ruffo S.}
\REVIEW{Phys. Rev. Lett.}{87}{2001}{030601}.

\bibitem{cohen}
\Name{Ispolatov I. \and Cohen E.G.D.}
\REVIEW{Physica A}{295}{2001}{475}.

\bibitem{ispo} 
\Name{Chavanis P.H. \and Ispolatov I.}
\REVIEW{Phys. Rev. E}{66}{2002}{036109}.

\bibitem{barreth}
\Name{Barr\'e J.}
\REVIEW{PhD Thesis}{\rm ENS-Lyon}{2003}{}.

\bibitem{at}
\Name{Antoni M. \and Torcini A.}
\REVIEW{Phys. Rev. E}{57}{1998}{R6233};
\Name{Torcini A. \and Antoni M.}
\REVIEW{Phys. Rev. E}{59}{1999}{2746}.

\bibitem{art}
\Name{Antoni M., Ruffo S., \and Torcini A.}
\REVIEW{Phys. Rev. E}{66}{2002}{R025103}.

\bibitem{gwl}
\Name{Griffiths  R.B., Weng C-Y , \and Langer J.S.}
\REVIEW{Phys. Rev.}{149}{1966}{301}.

\bibitem{binder}
\Name{Paul W., Heermann D.W., \and Binder K.}
\REVIEW{J. Phys. A: Math. Gen.}{22}{1989}{3325}.

\bibitem{rosti}
\Name{Rostiashvyyili V.G. \and Schilling R.}
\REVIEW{Z. Phys. B}{102}{1997}{117}.


\bibitem{ar}
\Name{Antoni M. \and Ruffo S.}
\REVIEW{Phys. Rev. E}{52}{1995}{2361};
\Name{Antoni M., Hinrichsen H. \and Ruffo S.}
\REVIEW{Chaos, Solitons \& Fractals}{13}{2002}{393}.

\bibitem{yama}
\Name{Yamaguchi Y.Y.}
\REVIEW{Phys. Rev. E}{68}{2003}{066210}.

\bibitem{muk}
\Name{Barr\'e J., Ruffo S. \and Mukamel D.} in \cite{leshouches}.

\bibitem{katzone}
\Name{Katz J.}
\REVIEW{Mon. Not. R. Astron. Soc.}{183}{1978}{765}.

\bibitem{katz}
\Name{Katz J. \and Okamoto I.}
\REVIEW{Mon. Not. R. Astron. Soc.}{317}{2000}{163}.

\bibitem{binney}
\Name{Binney J. \and Tremaine S.}
\Book{Galactic Dynamics}
\Publ{Princeton Univ. Press, Princeton}
\Year{1987}.

\bibitem{landau}
\Name{Landau L.D. \and Lifshitz E.M.}
\Book{Statistical Physics}
\Publ{Pergamon Press, Oxford}
\Year{1985}.

\bibitem{freddy}
\Name{Bouchet F.}
\REVIEW{preprint (cond-mat/030517)}{}{2003}{}.

\end{thebibliography}
\end{document}